\documentclass[sigconf]{acmart-edbt2023}

\def\BibTeX{{\rm B\kern-.05em{\sc i\kern-.025em b}\kern-.08em
    T\kern-.1667em\lower.7ex\hbox{E}\kern-.125emX}}

\usepackage{booktabs} 
\usepackage{graphicx}
\usepackage{subfig}
\usepackage{bbm}
\usepackage{color}
\usepackage{flushend}

\flushend

\newcommand{\argmax}{\operatornamewithlimits{argmax}}

\begin{document}
\title{Recommendation Systems in Libraries: an Application with Heterogeneous Data Sources}


\author{Alessandro Speciale}
\affiliation{%
  \institution{Politecnico di Torino}
}
\email{ale.speciale@studenti.polito.it}

\author{Greta Vallero}
\affiliation{%
  \institution{Politecnico di Torino}
}
\email{greta.vallero@polito.it}

\author{Luca Vassio}
\affiliation{%
  \institution{Politecnico di Torino}
}
\email{luca.vassio@polito.it}

\author{Marco Mellia}
\affiliation{%
  \institution{Politecnico di Torino}
}
\email{marco.mellia@polito.it}


 




\renewcommand{\shortauthors}{}

\begin{abstract}
The \emph{Reading$[\&]$Machine} project exploits the support of digitalization to increase the attractiveness of libraries and improve the users' experience. The project implements an application that helps the users in their decision-making process, providing recommendation system (RecSys)-generated lists of books the users might be interested in, and showing them through an interactive Virtual Reality (VR)-based Graphical User Interface (GUI).  
In this paper, we focus on the design and testing of the recommendation system, employing data about all users' loans over the past 9 years from the network of libraries located in Turin, Italy. In addition, we use data collected by the Anobii online social community of readers, who share their feedback and additional information about books they read. 
Armed with this heterogeneous data, we build and evaluate Content Based (CB) and Collaborative Filtering (CF) approaches. Our results show that the CF outperforms the CB approach, improving by up to 47\% the relevant recommendations provided to a reader. 
However, the performance of the CB approach is heavily dependent on the number of books the reader has already read, and it can work even better than CF for users with a large history. Finally, our evaluations highlight that the performances of both approaches are significantly improved if the system integrates and leverages the information from the Anobii dataset, which allows us to include more user readings (for CF) and richer book metadata (for CB). 
\end{abstract}

\maketitle

\section{Introduction}

In the last few years, the high penetration of Information and Communication Technology (ICT) has paved the way for the collection of a large amount of data regarding user behaviours and preferences. 
This has led to the design and deployment of services and applications able to meet the users' needs and improve their experience, in many sectors such as healthcare, commerce, agriculture and industries. Among this, the educational field is not an exception: with the new functionalities ICT offers, we have the opportunity to improve the services and offer a more personalised experience \cite{sirel2021reflection}.  
According to \cite{sirel2021reflection, csanli2022effects}, libraries shall meet the evolving digital revolution and user needs while preserving their traditional role of education, equality, transparency, and civilisation institutions at society's disposal. 

In line with this, the \emph{Reading$[\&]$Machine} project\footnote{\url{https://smartdata.polito.it/reading-machine/}} aims at implementing an application based on RecSys. The application is accessible by users through a VR GUI installed in the public libraries of Turin (Italy). Its goal is to help readers in searching and finding  the books that are most in line with their tastes. In this paper, we describe the preliminary phase of the project. It consists of the analysis of the datasets, their enrichment, and the design, implementation, and evaluation of different RecSys.  

In this work, we use two different data sources. The first provides a 9 years-long dataset, which collects the loans in all public libraries located in Turin (Italy); the second contains a 7 years-long dataset from the Anobii social network\footnote{\url{https://www.anobii.com/}} where an online community of more than a million reading enthusiasts share their reading experiences. With this data, we implement and evaluate a content-based (CB) and collaborative-filtering (CF) RecSys, using implicit user feedback. 
According to our evaluations, the CF performs better than the CB approach for our application. From the results, integrating the information from the Anobii dataset is effective to significantly improve both their performance: this allows to include more users, useful for the CF, and to enrich book metadata, useful for the CB. Finally, results highlight that the performances of CB are strictly dependent on the number of books already read by a user, while for the CF this history has almost no impact. 
 
The paper is organised as follows. Section \ref{sec:related_works} reviews related works, while in Section \ref{sec:databases} the used databases are described and characterised. The methodology and the Key Performance Indicators (KPIs) are presented in Sections \ref{sec:recsys} and \ref{sec:kpi}, respectively. Results are discussed in Section \ref{sec:results} and the conclusions are drawn in Section \ref{sec:conclusion}.

\section{Related Works}
\label{sec:related_works}
RecSys have applications in disparate domains of our daily-life. 
To weigh the users’ preferences, they need to have feedback from users. As claimed in \cite{nunez2012implicit,choi2012hybrid}, this information can be acquired explicitly, by collecting users’ ratings; or implicitly, by observing users’ actions \cite{wang2018personalized,rendle2012bpr}. Authors in \cite{bobadilla2013recommender} categorise the RecSys into two main families: CB and CF systems. CB systems focus on analysing user or item metadata only. As explained in \cite{salter2006cinemascreen,van2000using}, they make recommendations based on users' choices made in the past: given properties of items that the user likes, the system will suggest other items with similar properties. According to~\cite{candillier2007comparing,su2009survey}, CF systems make recommendations to each user based on information provided by those users we consider to have the most in common with them. In \cite{hsu2007aimed,christakou2007hybrid, staab2002intelligent}, authors propose the employment of RecSys for entertainment applications, to recommend TV programs, movies, and travels, respectively. In \cite{guo2007intelligent,teran2010fuzzy}, authors use RecSys to improve the effectiveness of e-government applications, while works presented in \cite{huang2004graph} use them for commercial services. 

To the best of our knowledge, in the literature there are few investigations on recommendation systems for books and libraries, likely due to the difficulties in getting extensive and reliable data.  
One of the few examples is the work discussed in \cite{zhang2016personalized}, where a CF approach is proposed for libraries used by college and university students. The students' selected courses and learning trajectories are taken into consideration to provide a timely recommendation. Differently, we focus on general-purpose libraries and compare multiple approaches.   
In \cite{srujan2018classification} the authors classify the textual comments given to books in Amazon, to determine whether the opinion is positive, negative, or neutral. However, the obtained classifier is not used in conjunction with a recommendation system.  

The Helsinki Central Library Oodi, in Finland, is an example of public library modernisation. 
Besides its sustainable and innovative architecture, it provides users with an Artificial Intelligence (AI)-based bot, called \emph{Obotti}, which implements a RecSys, to suggest books to users, according to what they read and their preferences \cite{hammais2017virtual}. It is based on six chatbots, each recommending specific content, that collect users' interests by chatting. Its RecSys addresses the challenge of \emph{choice overload} for library visitors, presented in \cite{iyengar2000choice,zhang2016personalized}. Indeed, while choosing, users face an enormous quantity of possibilities, which makes them confused. Differently from Obotti, in this paper we enrich library data with a social network dataset (Anobii), and we share with the research community the comparison of different recommendation systems.


\section{Datasets and characterization}
\label{sec:databases}

In this work, we leverage data provided by the public libraries located in Turin, Italy, called \emph{Biblioteche Civiche di Torino} (BCT). Then, we integrate them with data from the online social network for reading enthusiasts \emph{Anobii}. 

\subsubsection*{BCT Dataset} 
BCT is a network of 49 public libraries which hosts a large collection of books. Users can access these books through loans. 
The anonymized data useful for our goal are reported in the following tables:
\begin{enumerate}
    \item \emph{Books Table} contains the information of each distinct book that is present in the collection. In total there are $290\,125$ distinct books. Each has a unique identifier, called book ID, followed by the author(s), title,  type of the item (monograph, manuscript, DVD, etc.), and language of the edition. 
    \item \emph{Loans Table} contains details of the $5\,484\,078$ loans that occurred between 2012 and 2020. For each loan, i.e., for each row, the anonymized user ID is reported, as well as the date of the loan.  
    In total, we observe $163\,321$ users subscribed to the BCT that borrowed at least one book. The average number of loans per book is 5 (median is 4). 
    In addition, even if users borrow on average 33 items, 75\% of them has less than 24 items.
\end{enumerate}

For this work, we restrict our analysis to \emph{monographies} and \emph{manuscripts} written in Italian language (Books table), keeping $228\,059$ books for the analysis. 

\subsubsection*{Anobii dataset}
Born in 2006, Anobii is an online thematic platform specialized in books, where users can create their own virtual library, share opinions, and enter ratings and reviews. Acquired by Italian companies (Mondadori in 2014 and Ovolab in 2019), Anobii increased its popularity in Italy. Currently, it has over 1 million users worldwide, of which around $400\,000$ are in Italy. The following tables of Anobii dataset are relevant for this work:
\begin{enumerate}
    \item \emph{Items Table} contains the information of $8\,021\,517$ items, mostly books, which are discussed on the social network. In particular, for each book (identified by Item ID), we have author(s), title,  language, plot, and keywords. Moreover, each item is associated with genres, provided by users. As a result, each book has multiple genres, with the number of users who did this association reported. There are 41 possible genres and each book is associated with 4 genres on average. 
    \item \emph{Ratings Table} collects ratings of items as entered by users over time. Each row presents the anonymized User ID, the Item ID, and the ratings, which are expressed as integer values from 1 to 5, in increasing order of appreciation. In this work, we rely on ratings recorded from 2014 to 2021. We observe more than 52 million ratings from $1\,202\,909$ distinct users. The average number of ratings per book is~14, but 75\% of them are rated at most 7 times. 
    Similarly, the average number of ratings per user is 93 (median is~13).
\end{enumerate}

Similarly to what has been done for the BCT dataset, we focus on items that are books written in Italian. 
Moreover, in the Rating Table, we remove rows with ratings lower than 3, since we assume that those are negative feedback. The rationale is to recommend items that have received only positive feedback.

As mentioned above, multiple genres are associated with a book. We neglect genres associated with almost all books or with very few books (e.g, \emph{Fiction and Literature}, \emph{Textbooks}, \emph{References}, and \emph{Self Help}). To have the distribution of genres among books as balanced as possible, we aggregate some genres by considering the entropy value calculated using their occurrences, and the aggregation is performed if it leads to the entropy reduction. Finally, in order to make reliable the association of a genre to a book, we keep only the top 4 genres associations per book, according to the number of votes.  Each of the 4 genres has a  probability proportional to the number of association occurrences (the sum of the genre probability is equal to one for each book).

\begin{figure}[t]
    \centering
    \includegraphics[clip,width=\columnwidth]{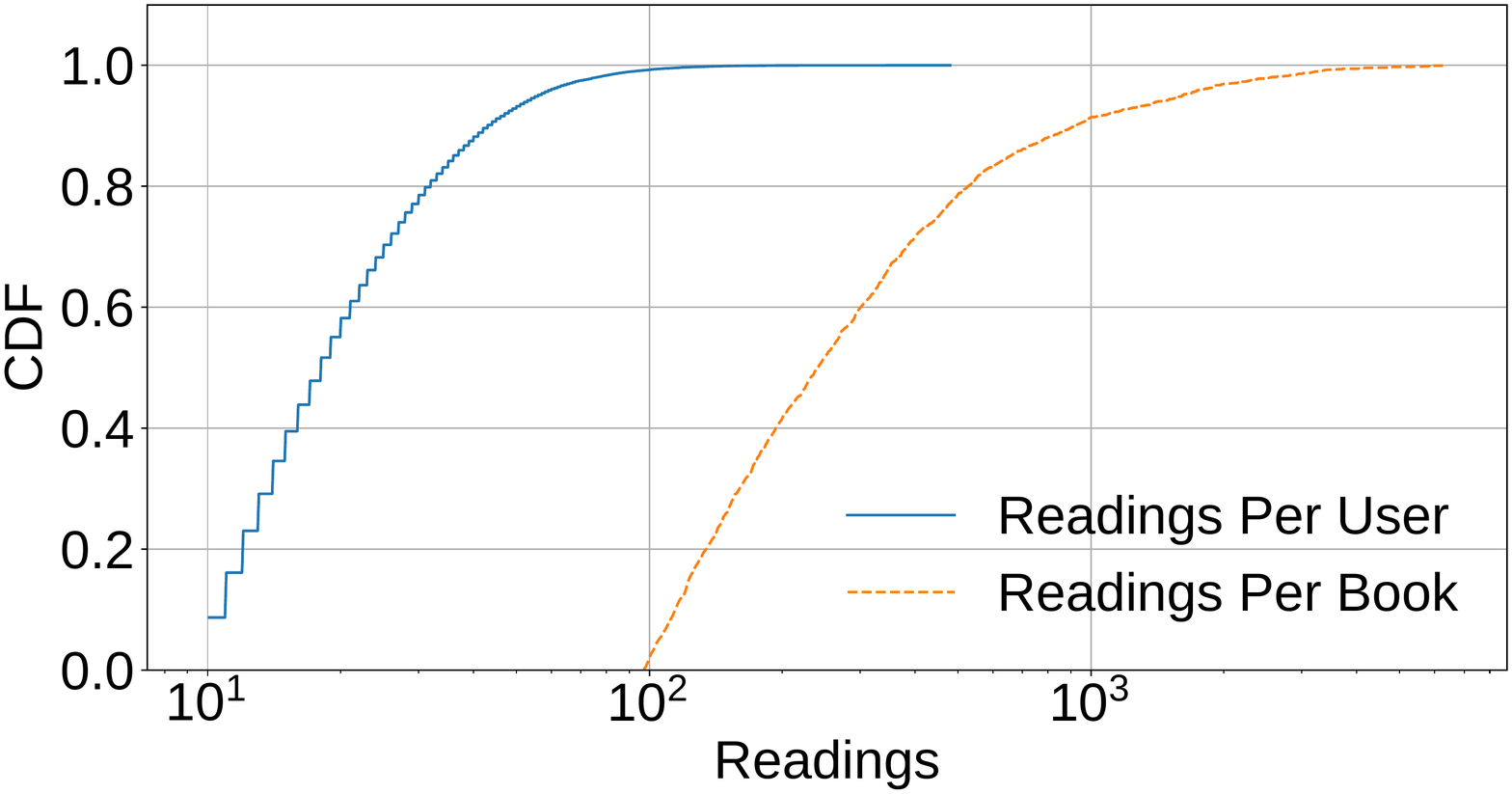}
    \caption{CDF of the readings per user and per book in the merged dataset, with a log-scale on x-axis.}
    \label{fig:cdfLoansND}
\end{figure}

\begin{figure}[t]
    \centering
    \includegraphics[clip,width=\columnwidth]{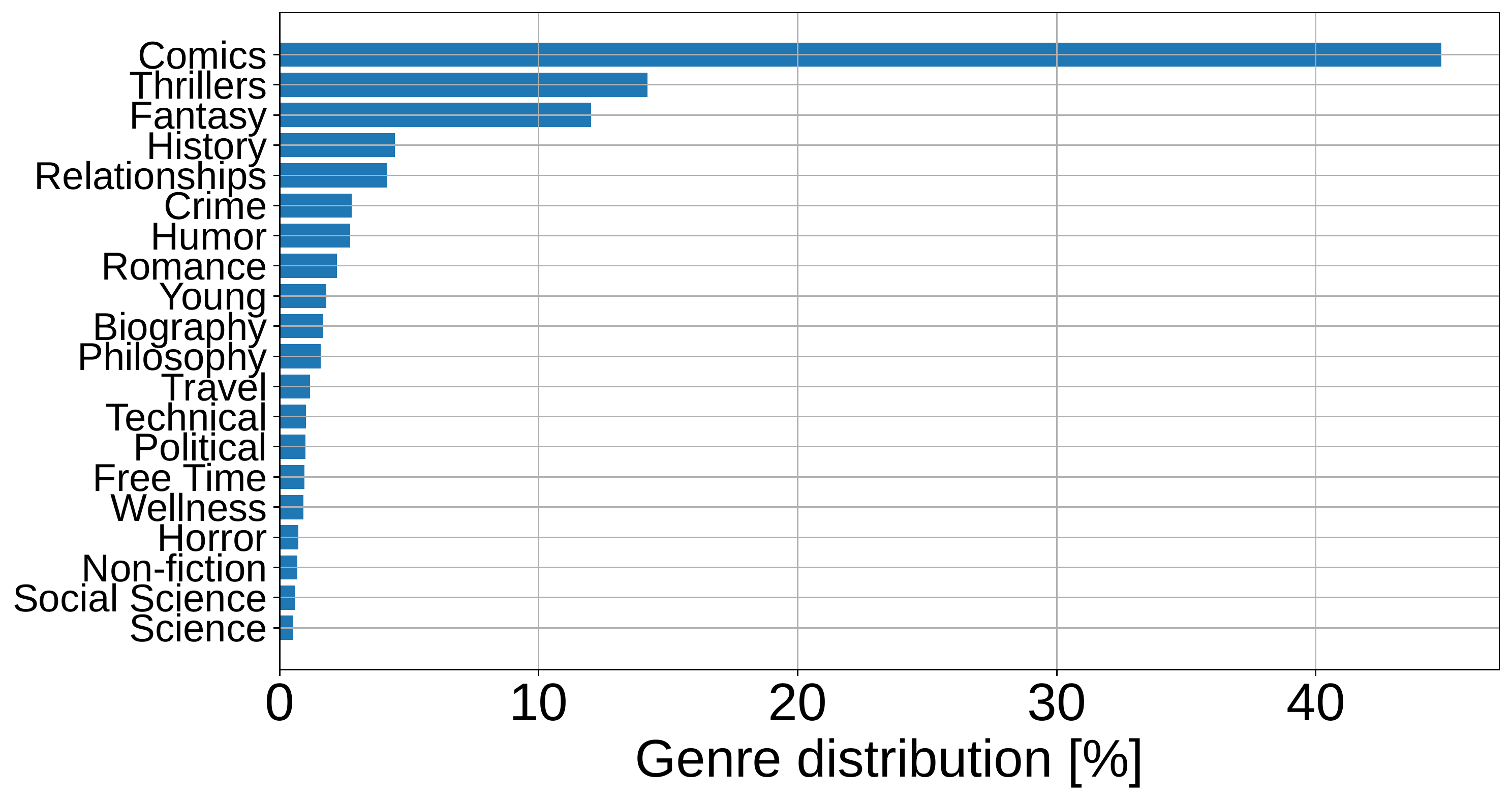}
    \caption{Distribution of the genres in the readings in the merged dataset.}
    \label{fig:genreDistribution}
\end{figure}

\subsubsection*{Merging BCT and Anobii datasets}

We generate a final dataset by joining the information from BCT and Anobii. 
For each book present in both the BCT and Anobii datasets, we keep all the attributes from both datasets that might be useful for the recommendation systems. 
We also create a \emph{Readings Table} that contains the Loans Table of the BCT dataset and the Ratings Table of the Anobii dataset, considering only the selected books. 
For this work, we drop users who read less than 10 books (around 86\% of them) and books which are read less than 100 times (around 89\% of them). We leave for future work the investigation of RecSys for users without a history or with a small one, as well as for niche books.
After this, we obtain $2\,332$ books and $43\,531$ users, $6\,079$ from BCT and $37\,452$ from Anobii. The loans are $1\,032\,277$. We report the CDF of the loans per user and per book in Fig. \ref{fig:cdfLoansND}.
We notice from the figure that the number of readings per user goes up to 480, while the number of readings per book goes up to~$6\,000$.

Next, we analyse the genre preferences. Each bar in  Fig. \ref{fig:genreDistribution} reports the percentage of how much each genre is read. 
It is possible to notice that among the loans, 44\% are relative to \emph{Comics}, followed by \emph{Thriller} and \emph{Fantasy}, which account for 14\% and 12\%, respectively. According to our data, users read multiple genres. Interestingly, 99\% of users read two genres at least ten times more than all the other genres togheter.   

\section{Recommendation Systems Design}
\label{sec:recsys}

In the case of explicit feedback, users quantify their appreciation through ratings, e.g., by giving between 1 and 5 stars. In this case, RecSys predicts the rating the user would give to an unread book and recommends the $k$ books with the highest ratings.  Unfortunately, user in libraries do not express any degree of appreciation for books, and we have only the list of their readings (implicit feedback). Therefore, we assume that if a user read a book, it is appreciated. Note that this is not always the case in practice: we leave for future work a study of possible features to reduce the limitations of this assumption, e.g., using the duration of the loan.
The RecSys provides a ranking to each user, which is a sorted list of books representing the possible level of appreciation, and recommends the top $k$ ranked books.
The value of $k$ is set in order to have a good trade-off between the quality of recommendations and the prevention of users' choice overload.

In this work, we consider different implicit feedback-based RecSys, which we detail in the following.

\subsubsection*{Random Items}
This approach is used as a baseline to understand if the RecSys is properly learning. Given a user $u$, it randomly recommends a set of $k$ books, which have not been read yet by that user.

\subsubsection*{Most Read Items}
This second baseline computes the number of times books are read in the training set, and then recommends the top $k$ most read books to all users. Therefore, the same recommendations apply to all users.

\subsubsection*{Closest Items}
This is a content-based RecSys. Its main idea is that users will read books similar to books they have already read in the past.  
In order to do this, given $N$, the books of the catalogue, we define $N_u$ as the set of books that user $u$ has already read,  and $\overline{N_u}$, given by $N \setminus N_u$,  the set of the books which the user $u$ dis not have read yet. For each book $b \in \overline{N_u}$ we compute~$s_b$, the average similarity to the books in $N_u$:
\begin{equation}
    {s_b} = \frac{\sum_{i\in N_u} s_{b,i}}{|N_u|}
    \label{ep:similarity}
\end{equation}
where $s_{b,i}$ is the similarity between book $b$ and book $i$ and $|N_u|$ is the cardinality of $N_u$. Once $s_b$ is computed for each $b$ in $\overline{N_u}$, we recommend the $k$ books with the highest $s_b$ to the user $u$.

For the computation of similarity $s_{b,i}$ we first extract the metadata of the books, to create a \emph{metadata summary}. 
It is a string given by the concatenation of the book's metadata. In this work we use all the possible combinations of (i) the book title, (ii) the author(s), (iii) the book plot, (iv) the genres, and (v) the book keywords.

To compute the similarity between the metadata summary of two books $s_{b,i}$, we need to derive the numerical representation of these strings as numerical vectors, that we use as embedding space. These vectors carry a semantic meaning with them, so that two items which are similar fall in the same region of the embedding space. 
We use Sentence Bidirectional Encoder Representations from Transformers (SBERT), which is a transformer for Natural Language Processing (NLP), developed by Google, see \cite{devlin2018bert}. For our implementation, we use the library described in \cite{reimers-2019-sentence-bert, reimers-2020-multilingual-sentence-bert}, which provides a pre-trained model, to map text given as input to the numerical embedding. Authors of \cite{reimers-2019-sentence-bert} trained SBERT using a labelled dataset of text pairs, for which the semantic distance is known. 
The semantic distance quantifies how much the two texts are similar (i.e., close in the vector space). 
The loss function aims at minimizing the error of the semantic distance prediction.  
Given the numerical representations of words, we compute  $s_{b,i}$ as the cosine similarity of the numerical representation of the metadata summary of the book $b$ and $i$, respectively. 

\subsubsection*{Bayesian Personalised Ranking (BPR)}
As presented in \cite{rendle2012bpr}, the BPR is a CF RecSys for implicit feedback. It adapts the Matrix Factorization (MF), introduced in \cite{rendle2010factorization}, to the implicit feedback case. 
Applying \cite{rendle2012bpr} to our scenario, given $U$ and $B$ the number of users and books, respectively, the user-item interactions matrix $I \in \mathbb{R}^{UxB}$ is the matrix where $i_{u,b}$ is 1 if user $i$ has read book $b$, 0 otherwise. Then BPR decomposes the matrix $I$ into the product of two lower-dimensional matrices, $V$ and $P$.
The first has a row for each user, while the second has a column for each book. The row associated with a specific user and the column associated with a specific book are called latent factors. The predicted user-item interactions matrix is calculated as $\tilde{I}=VP$, where $V \in \mathbb{R}^{UxL}$, given the number of users $U$ and latent factors $L$, and $P\in \mathbb{R}^{LxB}$, where $L$ is the number of latent factors and $B$ is the number of books.

We train the model, to learn the book's and user's latent factors which provide the rank of books for each user. The books read by a user are assumed to be preferred over unread books. Ranks are given according to a score $f(u,i|V,P)$ for a user $u$ and a book $i$, where V and P are the latent factor matrices we want to find. 
Therefore, we define the function: 

\begin{equation}
    p(i >_u j | V,P) = \sigma(f(u,i|V,P)-f(u,j|V,P))
    \label{eq:BPRprob}
\end{equation}

where $\sigma(\cdot)$ is the sigmoid function. Considering $N_u$ the set of read books for user $u$ and $\overline{N_u}$ the set of unread books for $u$,  we want to maximize this likelihood function $p(i >_u j | V,P)$ when $i\in N_u$ and $j \in \overline{N_u}$, i.e., read books should have a score larger than unread ones. 
Then, the latent factors are found through the following loglikelihood maximization over all users and pairs of read and unread books:

\begin{equation}
    \argmax_{V,P}  \sum_{u \in U} \sum_{i\in R_u, j \in \overline{R_u}} ln(p(i >_u j | V,P))-\lambda_{V} ||V||^2-\lambda_{P} ||P||^2
    \label{ep:BPRopt} 
\end{equation} 

where V and P are distributed according to a zero mean normal distribution with variance-covariance matrices obtained by multiplying the identity matrix with $\lambda_{V}$ and $\lambda_{P}$, respectively.  Therefore, $\lambda_{V} ||V||^2$ and $\lambda_{P} ||P||^2$ in Equation \eqref{ep:BPRopt} act as regularization terms. Further details and mathematical passages can be found in \cite{rendle2012bpr}.

To numerically perform the optimization in Equation \eqref{ep:BPRopt}, we use the variant of Weighted Approximate-Rank Pairwise (WARP) loss to learn model weights via Stochastic Gradient Descent, see~\cite{weston2011wsabie}. Given a read book $i\in N_u$, the WARP randomly takes an unread book $j \in \overline{N_u}$. If the score $f(u,i|V,P)$ exceeds $f(u,j|V,P)$ then the latent factors $V,P$ are updated. If this is not the case, another unread book $j \in \overline{N_u}$ is randomly picked. The magnitude of the update decreases with the number of extraction of unread books before the update, since unread books are extracted much more likely than read ones. Again, further details can be found in \cite{weston2011wsabie}.

\section{Key Performance Indicators}
\label{sec:kpi}

In this work, we measure the RecSys performances for BCT users, which are the target of the recommendation. We use 20\% of the readings of each BCT user as test set. The remaining part is further split into training and validation (80\% and 20\% of the remaining readings for each user, respectively). All the Anobii data are used for training (80\% of the readings of each user) and validation (20\%), without a test set. 

Consider $U$ users. Let $T_u$ be the books read by the user $u$ in the test set, and  $R_u$ the set of $k$ books recommended to the user~$u$. 
We quantify the performance with the KPIs described below, which depend on the choice of $k$. 

\subsubsection*{Number of Users with Relevant Recommendations (URR)}
Once the RecSys generates the recommendation for each user, we compute the fraction of users who have at least a recommended book in the test set:
\begin{equation}
    URR(k) = \frac{1}{U}\sum_{u}\mathbbm{1}_{ T_u \cap R_u \neq 	\emptyset}
    \label{eq:URR}
\end{equation}

where $\mathbbm{1}_{ T_u \cap R_u \neq 	\emptyset}$ is one if the intersection of the set of books of the test set read by the user $u$ and the recommended books is not empty. 

\subsubsection*{Average Number of Relevant Recommendations per Users (NRR)}
It accounts for the average number of books which are recommended by the RecSys and are in the test set:

\begin{equation}
    NRR(k) = \frac{1}{U}\sum_{u} |T_u \cap R_u|
    \label{eq:NRR}
\end{equation}

where $|T_u \cap R_u|$ is the cardinality of the intersection of the set of books of the test set read by the user $u$  and the set of books recommended to the user.

\subsubsection*{Precision (P)}
This metric quantifies the average ratio of books among the recommended ones, which are also in the test set. It is given by:

\begin{equation}
    P(k) = \frac{1}{U}\sum_{u} \frac{|T_u \cap R_u|}{|R_u|}
    \label{eq:P}
\end{equation}


\subsubsection*{Recall (R)}
It is the average ratio of books in the test set which are recommended:
\begin{equation}
    R(k) = \frac{1}{U}\sum_{u} \frac{|T_u \cap R_u|}{|T_u|}
    \label{eq:R}
\end{equation}

\subsubsection*{Average First Rank Position (FR)}
This evaluates the average rank of the first relevant recommendation over all the users. The rank of the first relevant recommendation for a user is the best position among the user's recommended books obtained by the books in the user's test set. A lower FR indicates a better performance. It does not depend on $k$.

\vspace{0.2cm}

Notice that all the proposed algorithms (Section \ref{sec:recsys}) and metrics are predicting and measuring relevant items that users read. However, we are not providing any serendipity to the users.

\section{Experimental Results}
\label{sec:results}

\begin{figure*}[t]
\centering
    \subfloat[URR and NRR]{\includegraphics[
    clip, width=\columnwidth]{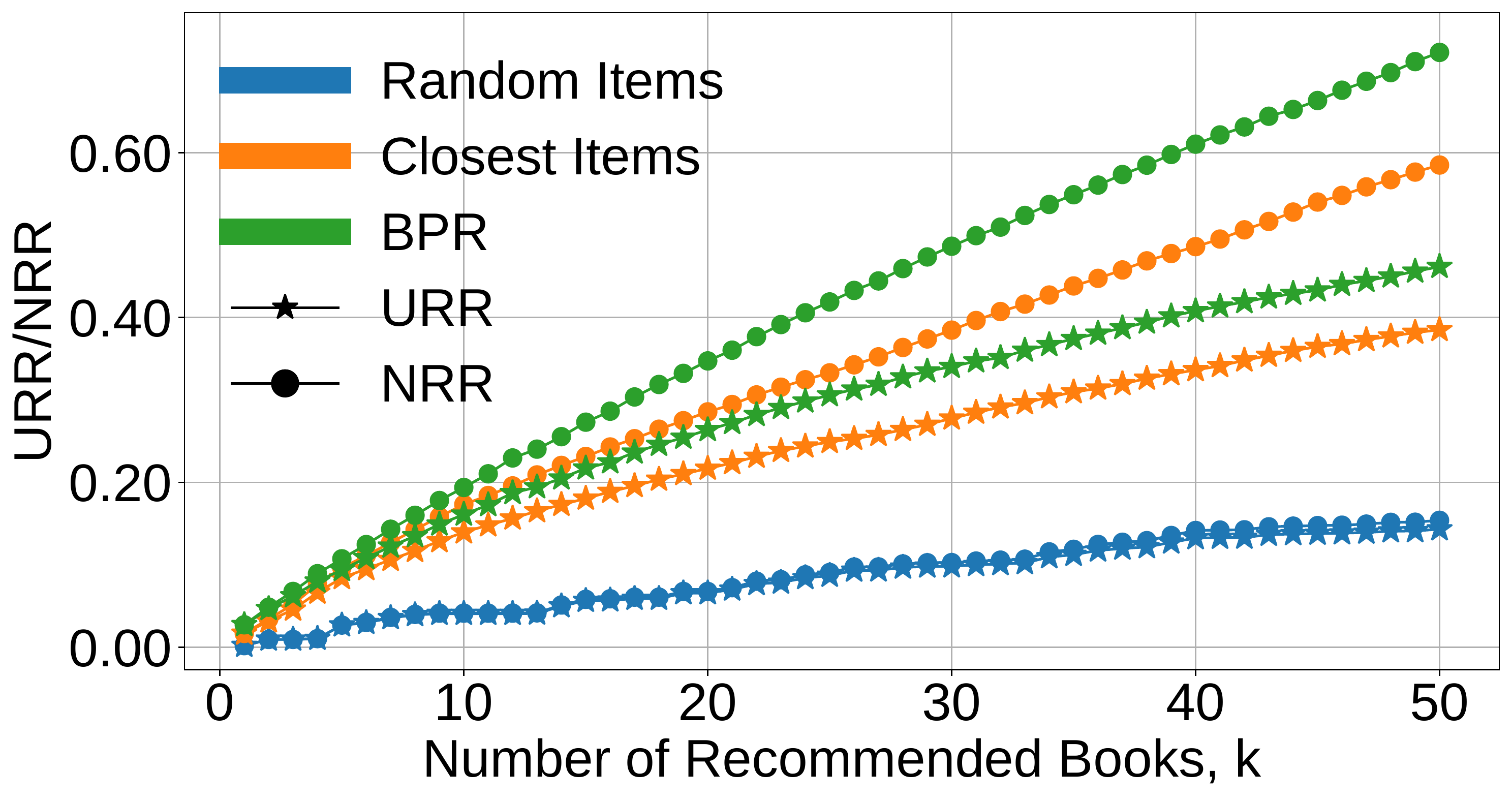}
    \label{fig:kvarURRNRR}}
    \subfloat[P and R]{\includegraphics[
    clip, width=\columnwidth]{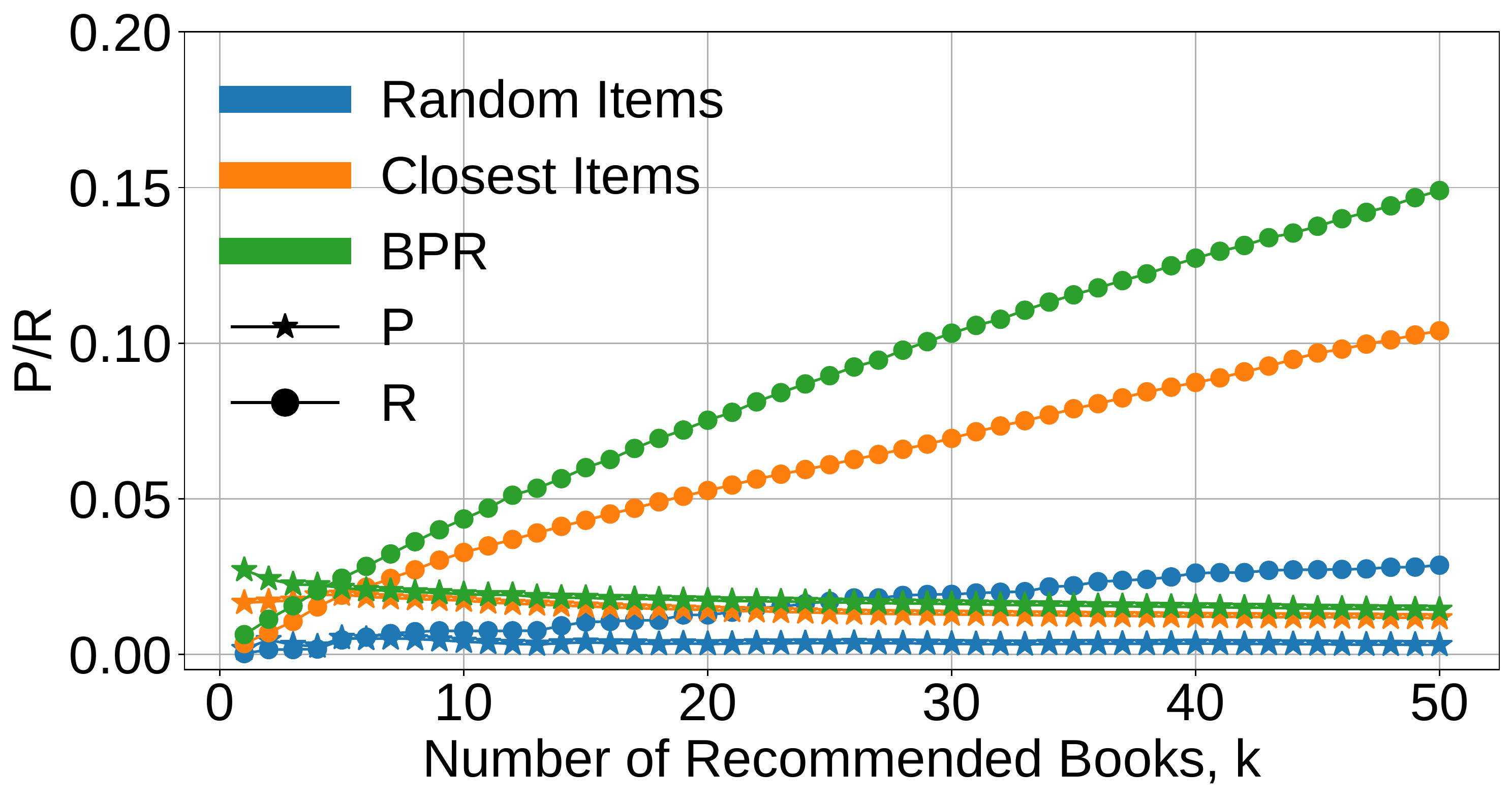}
    \label{fig:kvarRP}}
\caption{Effects on URR and NRR (a), P and R (b), varying the number of recommended books $k$.}
\label{fig:resultsKVar}
\end{figure*}

\begin{table}[]
    \caption{Results of the different RecSys with $k$=20} 
    \small

    \begin{tabular}{|l|l|l|l|l|l|} \hline
                     & URR  & NRR      & P    & R   & FR  \\ \hline \hline
    Random Items & 0.07 & 0.07 & 0.00 & 0.01 & 370 \\ \hline
    Most Read Items &  0.03 & 0.03 & 0.00 & 0.01 & 556 \\ \hline
    Closest Items & 0.22 & 0.29 & 0.01 & 0.05 & 186 \\ \hline
    BPR & 0.26 & 0.35 & 0.02 & 0.08 & 130 \\ \hline
    BPR (BCT only) & 0.15 & 0.17 & 0.01 & 0.04 & 298 \\ \hline
    \end{tabular}
    \label{tab:results}
\end{table}


In this section, we discuss the results with the different RecSys explained in Section \ref{sec:recsys}. 

First, we perform a grid search for the parameters of the BPR model. In particular, we vary the number of latent factors $L$ and the learning rate, which is the magnitude of the update of the latent factors $V,P$, at each iteration of the training phase. 
We choose those that maximize the URR on the validation set. The results reveal that $20$ latent factors and $0.2$ as learning rate provide the best performance.
For the Closest Items, the \emph{metadata summary} is built concatenating the author(s) and the genre(s) of the book, which is the best parameters combination (see Section~\ref{sec:metadata}).


Table \ref{tab:results} details the values for the KPIs with 20 recommended books (i.e., $k=$20). This is the value we choose in our application since it is a good trade-off between the quality of recommendations and the prevention of users' choice overload. 
As expected, non-personalized RecSys such as Random Items and Most Read Items perform poorly. The BPR algorithm obtains the best performance, outperforming the Closest Items by up to 47\%, providing the highest URR, NRR, R, and P. 

Notice that the obtained values of the KPIs seem low because we have to choose only $20$ books over $2\,332$ available ones, where only a small portion are actually read (the median of the readings per user is 18). In fact, the results reveal that both Closest Items and BPR provide significantly better performance than the Random Items and Most Read Items approaches. 

In the Table, we also report results obtained when BPR is trained using users from BCT only, denoted as $BPR (BCT only)$. The lower results with respect to the BPR case highlight the importance of integrating the Anobii dataset for obtaining good results. Given the poor performance, for the remainder of the paper, we do not show any more results for BPR trained only on BCT data and also for Most Read Books.


Next, we compare results obtained with the different RecSys by varying $k$. Fig. \ref{fig:kvarURRNRR} shows the URR and NRR, while Fig. \ref{fig:kvarRP} reports Recall and Precision, marked by stars and circles, respectively. In both the figures, we vary the number of recommended books $k$ between 1 and 50, for the Random Items (in blue), Closest Items (in orange), and BPR (in green). 
As expected, the URR, NRR, and R grow with the number of recommended books. Indeed, when the number of recommended books $k$ grows, having more relevant recommendations is more likely. However, P decreases since it considers the ratio of relevant recommendations divided by the number of recommended books.

\subsection{Impact of the Number of Read Books}
\begin{figure}
    \centering
    \includegraphics[
    clip, width=\columnwidth]{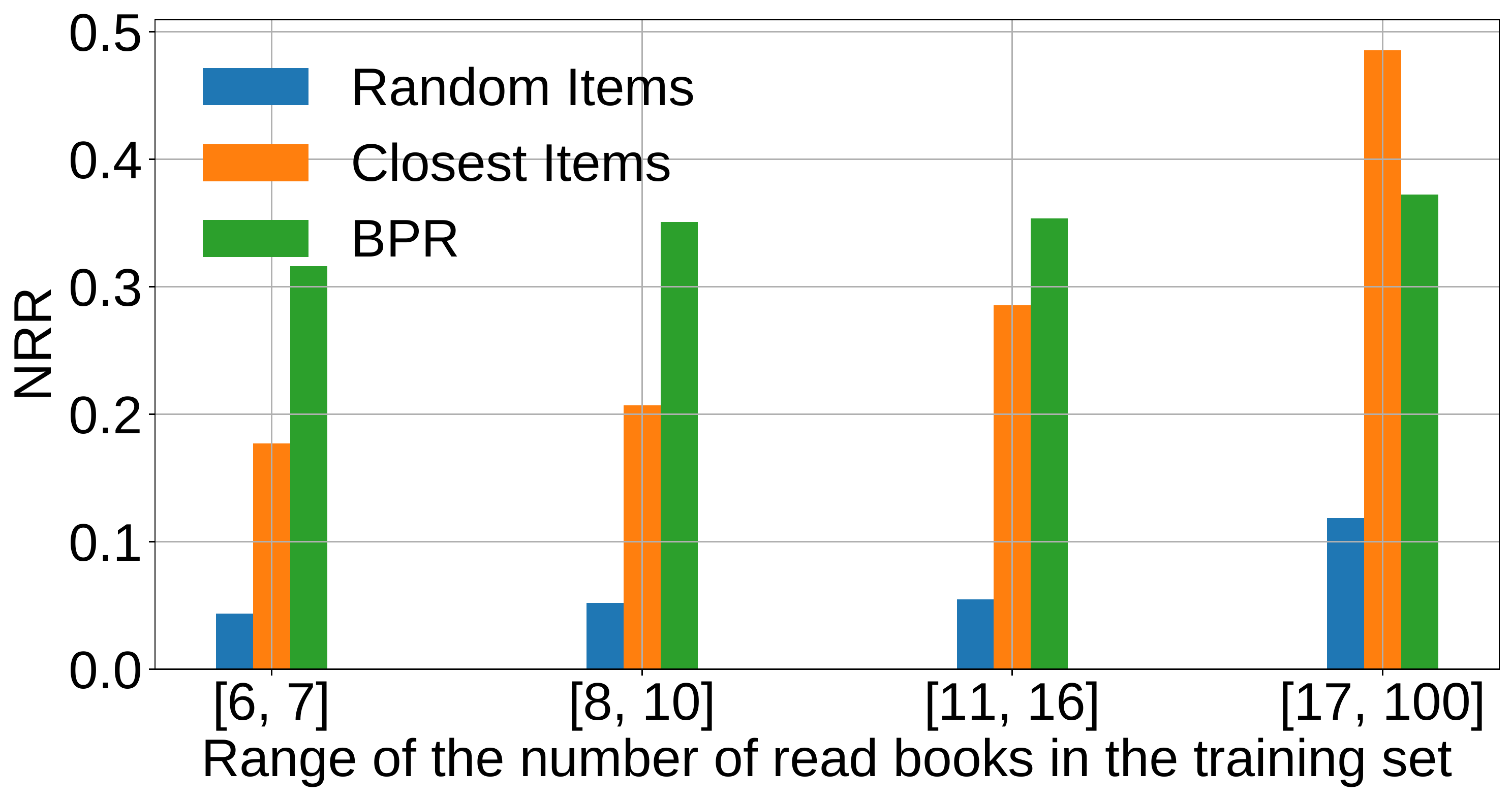}
    \caption{NRR on the test set, varying the number of books per user in the training set, with the number of recommended books $k$ equal to 20.}
    \label{fig:booksVariation}
\end{figure}

We analyse the performance of the trained recommendation systems on different groups of users. In particular, we analyse whether and how the results change for users that read a different number of books, i.e., users that have a different number of books that belong to the training set (recall that, for each user, we select 80\% of books for training and validation, and test recommendation on the remaining 20\% of them).
Fig. \ref{fig:booksVariation} reports the NRR with Random, Closest and BPR algorithms, in blue, orange, and green, respectively. 
The interval bins are chosen to have approximately the same number of users in each group.

We clearly notice  that the growth of the number of books read by users improves the NRR, with all used algorithms. 
This is expected since the probability of recommending a book that belongs to the set of read books increases with the size of the latter set. The growth of the NRR for the Random RecSys testifies this. Moreover, when many books are available in the training set, the preferences of a user can be better caught. 
For the users who have a number of readings in the training set that is lower than 8 
and between 8 and 10, the BPR obtains NRR equal to $0.32$ and $0.35$, respectively,  while Closest Items $0.18$ and $0.21$. 
Nevertheless, the growth of the users' readings has a small impact on BPR. Conversely, the Closest Items approach benefits by a sizeable increase in the NRR, reaching $0.29$ and $0.49$ when the number of readings in the training set ranges from 11 to 16 and from 17 to 100, respectively. This last case is 35\% better than the BPR case. For BPR, the effect of the growth of the number of readings for a user is less evident being a CF, and even a few readings are sufficient to exploit the user preferences of similar users. 

\subsection{Impact of the Metadata Summary}
\label{sec:metadata}

\begin{figure}
    \centering
    \includegraphics[
    clip, width=\columnwidth]{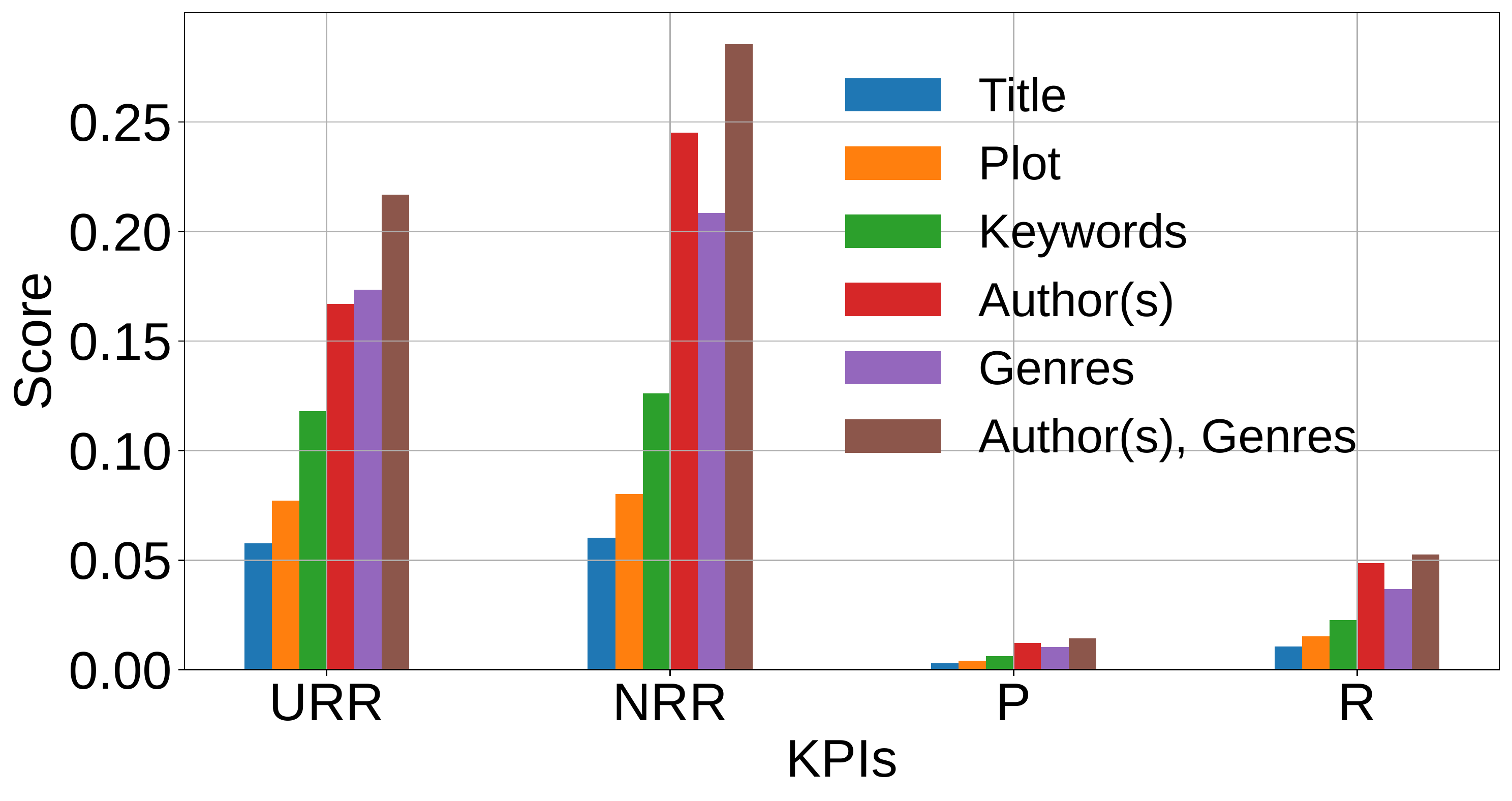}
    \caption{KPIs varying the composition of the book metadata summary for computing Closest Items, with the number of recommended books $k$ equal to 20.}
    \label{fig:metadataImpact}
\end{figure}

For the Closest Items approach,  we investigate the impact of the different \emph{metadata summary}, to compute the distance between books. 
In Fig. \ref{fig:metadataImpact} we show the different KPIs, obtained when the \emph{metadata summary} is composed of different concatenations of metadata (we do not show all the possible combinations, for lack of space). 
Using only the title as metadata (blue bar) results in the worst performance, similar to the Random approach. This suggests that users do not read books with titles similar to those already read. Already using the book plot or keywords (orange and green bars) to build the \emph{metadata summary} provides better performance. 
The author(s) alone (red bar) significantly improves the results, suggesting that many users like to read books from the same author. Results further improve by using the genres of the book (purple bar) on some KPIs, implying that users are attracted to books with genres similar to the ones already read. 
Among all the combinations, we obtain the best results using only authors and genres (brown bar). 
Not shown in the figure, we observe that adding the keywords to the best combination, i.e., concatenating them to authors and genres, slightly decreases the overall performance.



These results confirm the importance of the integration of the Anobii dataset for obtaining good results, emerged in section \ref{sec:results}. Indeed, Anobii provides books' metadata such as genres and keywords that are not available in the original BCT dataset.

\subsection{Training and Recommendation Time}

Finally, we focus on the running time needed for the training and the recommendation phase, 
that we report in Table \ref{tab:time}. The Random and Closest algorithms do not have a proper training phase, while, as indicated in the table, the BPR algorithm needs $30.55$~seconds with our dataset.

For the recommendation phase, we average over users the time which elapses between the reception of a request for a recommendation and the generation of the recommendations. 
Table~\ref{tab:time} reports it for each algorithm. It is evident that BPR requires a little more time than the others, but it is still acceptable for the application.

\begin{table}[]
    \caption{Average time in seconds needed for the training and recommendation generation phases.}
    \small
    \begin{tabular}{|l|l|l|} \hline
                     Time needed for: & Training (s) & Recommendation (s)     \\ \hline \hline
    Random Items     & - & 0.04  \\ \hline
    Closest Items    & - & 0.04  \\ \hline
    BPR              &  30.55 & 0.05   \\ \hline
    \end{tabular}
    \label{tab:time}
\end{table}

\section{Conclusion}
\label{sec:conclusion}
In this paper, we discuss the first steps of the \emph{Reading$[\&]$Machine} project, aiming at increasing the attractiveness of public libraries by designing an application capable of suggesting books tailored to the preferences of the readers.
We combine data from loans in libraries and from evaluations and characteristics of books crowdsourced within a social network (Anobii). We consider a CB and a CF approach, and results show that the CF increases the number of relevant recommendations provided to a user by the CB system, whose performance is strictly dependent on the number of books that the user has already read. 
For our application, the integration of the Anobii dataset, which includes more users and books' metadata, useful for the CF and CB, respectively, significantly improves the performances. 

It is important to note that the metrics defined in Section \ref{sec:kpi} are objectively trying to predict the next relevant books that users read. However, it would be interesting for future work also to consider books that are not related to the ones already read, but would be liked by the users, introducing parameters and metrics for evaluating the diversity and serendipity of the recommendations, as well as the possible boredom effect \cite{reccBoredom}.
Moreover, we do not take into consideration the user specific sequence of loans, namely the fact that a book has been chosen after another. Therefore, we could consider sequential recommendation systems algorithms (see~\cite{ijcai2019p0883}).

\begin{acks}
This work has been supported by the `FacciamolaFacile' grant funded by Fondazione TIM for the project `Reading (\&) Machine', and by the 
 `National Centre for HPC, Big Data and Quantum Computing' (CN00000013, Bando M42C, D.D. n. 3138, Decreto del MUR n. 1031).
\end{acks}
\bibliographystyle{ACM-Reference-Format}
\bibliography{biblio}

%
% ...

\end{document}